\documentclass[11pt,twoside]{article}
\usepackage{./asp2014}
\usepackage{longtable}
\usepackage{float}

\usepackage{array}
\newcolumntype{L}[1]{>{\raggedright\let\newline\\\arraybackslash\hspace{0pt}}m{#1}}
\newcolumntype{C}[1]{>{\centering\let\newline\\\arraybackslash\hspace{0pt}}m{#1}}
\newcolumntype{R}[1]{>{\raggedleft\let\newline\\\arraybackslash\hspace{0pt}}m{#1}}

\aspSuppressVolSlug
\resetcounters

\begin{document}

\title{Reaching Communities and Creating New Opportunities with the ngVLA}
\author{Lyndele von Schill and Suzanne Gurton}
\affil{National Radio Astronomy Observatory, Charlottesville, VA USA; \email{lvonschi@nrao.edu}, \email{sgurton@nrao.edu}}

\paperauthor{Lyndele vonSchill}{lvonschi@nrao.edu}{ORCID_Or_Blank}{NRAO}{ODI}{Charlottesville}{VA}{22903}{USA}
\paperauthor{Suzanne Gurton}{sgurton@nrao.edu}{ORCID_Or_Blank}{NRAO}{EPO}{Charlottesville}{VA}{22903}{USA}

\begin{abstract}
The Office of Diversity and Inclusion (ODI) and the Education and Public Outreach (EPO) Department serve the strategic goal of the National Radio Astronomy Observatory (NRAO) to broaden public awareness of, support for, and participation in Science, Technology, Engineering, and Mathematics (STEM). ODI operates a suite of programs designed to support underrepresented minority undergraduate, graduate students in pursuit of careers in STEM. EPO highlights the discoveries, technologies, and careers pioneered and exemplified by the NRAO via multipurpose engagement strategies that include face-to-face and standalone learning programs, products, and public services for the general public and K-12 students, with attention to reaching diverse audiences. These established and diverse programs are described, along with proposals for new, unique opportunities enabled by the development and realization of a next-generation Very Large Array (ngVLA).
\end{abstract}

\section{ Broadening Participation}
As a facility under the management of the National Radio Astronomy Observatory (NRAO), the next-generation Very Large Array (ngVLA) has full access to NRAO's suite of widely well-regarded Broadening Participation programs. 
NRAO's programs address the key priorities identified by the National Science Foundation (NSF) that include:
\begin{itemize}
\item Preparing a diverse, globally engaged science, technology, engineering, and mathematics (STEM) workforce; 
\item Integrating research with education, and building capacity; 
\item Expanding efforts to broaden participation from underrepresented groups and diverse institutions across all geographical regions in all NSF activities; and 
\item Improving processes to recruit and select highly qualified reviewers and panelists.
\end{itemize}

\begin{figure}[!ht]
\centering
\includegraphics[width=0.7\textwidth]{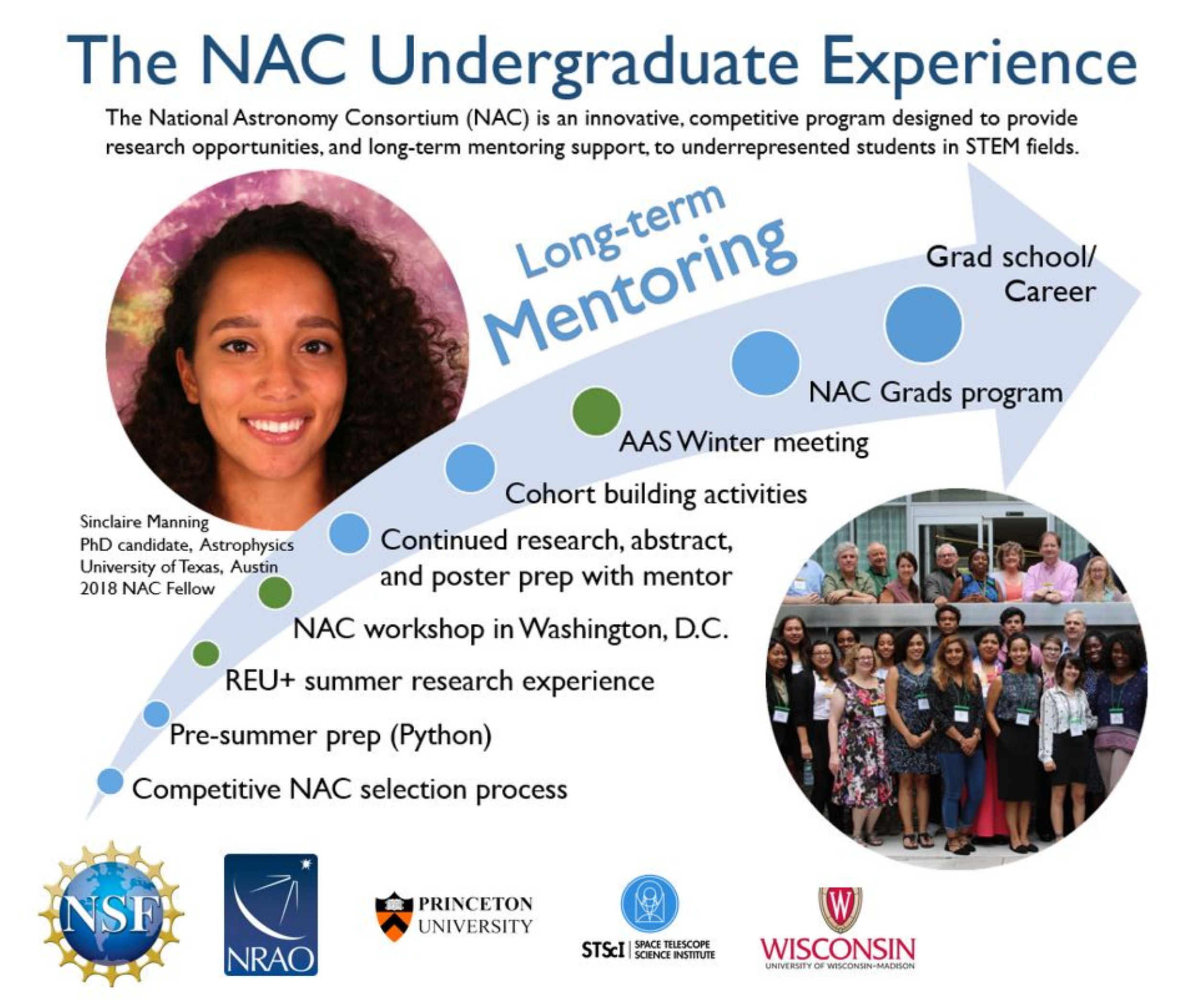}
\caption{\label{fig:BI-pipeline}   
The NRAO's Broadening Participation ``Pipeline".  }
\end{figure}

The NRAO has adopted a comprehensive, observatory-wide, "pipeline" approach to the development of STEM capacity, with particular emphasis on broadening participation of underrepresented groups and diverse institutions. This approach begins with engaging K-12 students in a seamless flow of STEM activities that relate to the full spectrum of fields associated with radio astronomy, and continues through providing undergraduate, graduate, and post-doctoral research, training, education, and mentoring programs (described below). The ngVLA will vastly increase the opportunities that the NRAO can offer to the U.S. and international public, student, and scientific communities.

In addition to its "outward-facing" broadening participation efforts, the NRAO has developed a set of policies and practices that are designed to recruit and select highly qualified, and diverse, reviewers, panelists, and search committees, with the goal of ensuring that selections are fair and equitable across all populations. 

Examples of the NRAO's existing, significant investment in broadening participation efforts appear in this section (see Figure \ref{fig:BI-pipeline}). Discussions about new opportunities generated by the ngVLA appear in Section 4 (Emerging Opportunities for Broader Impact and Broader Participation through the VLA).\\

\noindent 
\textbf{K-12 Education Programs:} 
	The NRAO's education programs are designed to introduce the hidden universe revealed by radio astronomy by engaging program participants in age appropriate inquiry based activities and research. Examples of K-12 education programs aimed at broadening participation include:
  
\begin{itemize}
\item \textbf{Radio Astronomy and Physics in New Mexico (RAP-NM):} A one-week residential summer camp experience on the NM Tech campus for rising 9$^\mathrm{th}$ graders from around the state of New Mexico. Student mentors are first recruited from the National Astronomy Consortium (NAC)  program (see below).
\item \textbf{Sister Cities and Observatories:} A 10-day international exchange recruiting from high school youth in communities near the Atacama Large Millimeter/submillimeter Array (ALMA) and VLA introducing them to the world class observatories and the diverse range of careers possible at each.  This program will continue under ngVLA.
\end{itemize}

\noindent 
\textbf{National Astronomy Consortium (NAC) Program:}
Opportunities for undergraduate research will be coordinated through the NRAO's NAC program, which is designed to provide research opportunities to underrepresented minority (URM) students. Students are recruited for participation in this program from an established network of Historically Black Colleges and Universities (HBCUs) and Hispanic-serving Institutions (HSIs). The ngVLA will open more opportunities for traditionally underrepresented students to participate in cutting-edge research which, in turn, increases opportunities for the students to attend graduate school and/or begin careers in radio astronomy.
\\

\noindent 
\textbf{National and International Non-Traditional Exchange (NINE) Program:}  
Collaborative opportunities also exist with the NRAO NINE, which trains the next generation of scientists and engineers from countries in which radio astronomy expertise is limited, but needed. The NINE program focuses on training traditionally underrepresented populations in skills that result in meaningful contributions to the astronomical science body of knowledge. The NINE program includes a network of ever-growing national and international "Hubs" where regional populations learn state-of-the art science and technology relevant to astronomy.  The ngVLA will provide cutting edge research opportunities to the NINE program, while benefiting from access to skilled scientists and technicians from the broad network of NINE Hub communities.
\\

\section{Public Engagement}
\noindent 
\textbf{Public Website:}  
In addition to programs designed with face-to-face interactions, the ngVLA will have a strong presence on the NRAO website that will serve as a portal for the public to explore the engineering advances and astronomical discoveries made possible by ngVLA. Clear information for visitors wishing to visit the site will be on the website, as well as opportunities to explore the site virtually. An ngVLA Explorer\footnote{e.g., \url{https://public.nrao.edu/special-features/vla-explorer/}}, combining video and augmented reality, will be created to give those who cannot visit the facility a virtual tour. The Role Model\footnote{\url{https://public.nrao.edu/special-features/role-models/}} series will be expanded to include ngVLA staff. It currently reflects the diversity of jobs and experiences that are needed to run a national observatory.  
\\
 
\noindent 
\textbf{Visitor Center:}  
The planned upgrade of the VLA Visitor Center (VC) reflects modern interpretive methods to explore the intersection of three realms: The stories arising from the resource, visitors' intrinsic interests, and the mission and goals of NRAO.  The goals for the ngVLA VC  will echo those for the VLA VC, with a substantial update to be inclusive of the international collaborations that will help make the ngVLA possible. These interpretive goals will be expressed through an overarching theme that the ngVLA serves humanity's deep curiosity and drive to explore the universe and our relationship to it:

\begin{itemize}
\item Interpretation will make information understandable and relevant for non-scientists as well as scientists.
	\begin{itemize}
	\item Non-scientist visitors will understand in general terms the basics of radio astronomy and the ngVLA - and feel happily surprised that they can.
	\item Scientists will appreciate the clarity of the message, without over-simplification to the point of introducing inaccuracies.
	\end{itemize}

\item  People of diverse backgrounds will feel welcome and actively included as the VC will actively: 
	\begin{itemize}
	\item Provide a venue for diverse cultural voices.
	\item Include women, persons of color.
	\item Encourage Navajo and Puebloans to share their stories.
	\end{itemize}

\item Educators will value and use the ngVLA as a resource.

\item Visitors and residents will understand how natural conditions here support the VLA, and will feel inspired to help maintain those conditions.

\item Visitors will understand why the ngVLA exists, and feel the program is useful, relevant and worth supporting.  
	\begin{itemize}
	\item People will understand how the ngVLA fits into the bigger picture of astronomy and human discovery.  
	\item Users will feel they are an integral part of the universe, not separate from it.  
	\item Local residents and visitors will feel pride and ownership in the ngVLA's role in unlocking the secretes of the universe.  
	\end{itemize}

\item Visitors will share their positive experiences with others back home.  
\end{itemize}

\noindent 
\textbf{Media Relations}
NRAO's EPO department includes a full media relations team with public information officers, artists and graphic designers to best represent the discoveries made possible by ngVLA  observations in the popular science media. News and information about ngVLA will be featured in such national outreach venues as the U.S.A. Science and Engineering Festival and National Astronomy Night on the Mall. 

\section{Broader Impacts: Training the Next Generation}
In addition to the NAC and NINE programs that focus on providing research, engineering, and other "full spectrum" astronomy fields, to underrepresented students, the NRAO has provided substantial internship opportunities for undergraduate and graduate students through its REU, graduate research, and NM Ops undergraduate internship programs. The NM Ops internship program, for example, has provided engineering students experience by working in the Observatory's electronics lab. The ngVLA will provide an opportunity to significantly increase the number of students with access to state-of-the-art research and engineering projects. 

For the last 2 years EPO has funded a media intern at NRAO headquarters each summer, with the inauguration of ngVLA a media intern will also be stationed at the ngVLA to supplement the efforts of the Public Information Officer and assist with major media efforts.

\section{Emerging Opportunities for Broader Impact and Broader Participation through the ngVLA}

The NRAO recognizes that the ngVLA project will span more than 10 years; this longevity offers a unique opportunity for the Observatory to build a pipeline for future employees from the regions in which the ngVLA will be present. Our BI and BP plans, then, include a proactive, intentional plan to include workforce development as an integral part of the entire ngVLA project, from beginning construction through and beyond first science. We expect to continue to ask the community for input into this important component of the ngVLA project; we have, however, identified the following opportunities for new, substantial, outcome-oriented Broader Impact (BI) and Broadening Participation (BP) activities (see Table~\ref{tbl:BI}). 

\begin{table}[!t]
\caption{Potential Broader Impact/Participation Opportunities Enabled by the ngVLA}
\label{tbl:BI}
\begin{center}
\begin{tabular}{|L{8.0cm}|C{1.0cm}|L{3.0cm}|}
\tableline
\noalign{\smallskip}
{\bf Emerging Opportunities through ngVLA} & {\bf BI/BP} & {\bf Communities Impacted}\\
\noalign{\smallskip}
\tableline
\tableline
\noalign{\smallskip}
Enhancement of broadband Internet access infrastructure, including fiber optics, as an intentional by-product of ngVLA internet needs.  & BI & small, rural communities along antenna paths\\

\hline
Enhancement of emergency services (e.g., fire response units) through collaboration with local communities. & BI & small, rural communities along antenna paths\\

\hline
Educational opportunities to engage and inform local stakeholders about radio astronomy and the importance of `their backyard' to important national/international science.  
\begin{itemize}
\vspace{-6pt}
\item Educational sessions will be presented by a diverse group of students and post-doc to raise visibility of both radio astronomy and diversity in STEM.
\vspace{-6pt}
\item Role model videos for grade school girls and URM students in all fields that support an Observatory.
\end{itemize} & BI \& BP & small, rural communities along antenna paths\\

\hline
Opportunity for partnership with URM college or university to develop educational and historical films:
\begin{itemize}
\vspace{-6pt}
\item Filming of the ngVLA build  for EPO and historical reference.
\vspace{-6pt}
\item Filming at each site to document community stories and engagement with the ngVLA.
\vspace{-6pt}
\item K-12 lesson plans developed to accompany films and film segments.
\end{itemize} & BI \& BP & K-12, Undergraduate and graduate students; faculty; and members of underrepresented STEM groups (e.g., African-American, Hispanic, and Native American students and faculty); the general public\\
\hline

\end{tabular}
\end{center}
\end{table}

\setcounter{table}{0}
\begin{table}[t]
\caption{Continued}
\begin{center}
\begin{tabular}{|L{8.0cm}|C{1.0cm}|L{3.0cm}|}
\noalign{\smallskip}
\hline
Opportunities for collaborative, cross-discipline research during archaeological site excavations during the pre-construction period:
\begin{itemize}
\vspace{-6pt}
\item Research opportunities with multiple departments (e.g., archeology/environmental sciences, water quality) at community colleges and universities (e.g., Dry Lake Bed).
\vspace{-6pt}
\item Efforts will be made to identify URM students for participation in the research opportunities.
\vspace{-6pt}
\item Development of lesson plans centered around the archeology and environment of ngVLA sites.
\vspace{-6pt}
\item Educational material produced for local communities and the general public, with a focus on connection to the land on which ngVLA antennas are located.
\end{itemize} & BI \& BP & K-12, Undergraduate and graduate students; faculty; and members of underrepresented STEM groups (e.g., African-American, Hispanic, and Native American students and faculty); the general public; the scientific community\\

\hline
Opportunities to incorporate non-Western cosmologies into the education about the ngVLA:
\begin{itemize}
\vspace{-6pt}
\item Invite indigenous community members (expert and local) to organized talks about indigenous cosmology.
\vspace{-6pt}
\item STEAM projects (e.g., murals, stories, videos) to capture non-Western cosmologies in local schools and community centers.
\vspace{-6pt}
\item Collaborate with the Society for Advancement of Chicanos/Hispanics and Native Americans in Science (SACNAS) to engage experts and students in these projects, with a plan to publish comparisons of western and indigenous cosmologies in the context of the ngVLA locations.
\vspace{-6pt}
\item Development and delivery of Star Parties that incorporate different cosmologies.
\end{itemize} & BI \& BP & Local community members; the general public; K-12, undergrad and graduate students; post-docs and faculty; the scientific community\\
\hline
Investigate additional benefits/applications of radio frequency interference (RFI) mitigation:
\begin{itemize}
\vspace{-6pt}
\item Providing research opportunities to URM students.
\vspace{-6pt}
\item Publications.
\vspace{-6pt}
\item Inventions.
\end{itemize} & BI \& BP & General public; industry; local communities; K-12, undergrad and grad students; faculty; scientific community\\

\hline

\end{tabular}
\end{center}
\end{table}

\setcounter{table}{0}
\begin{table}[t]
\caption{Continued}
\begin{center}
\begin{tabular}{|L{8.0cm}|C{1.0cm}|L{3.0cm}|}
\noalign{\smallskip}
\hline
Develop ``road shows" to provide information -- along the project timeline --  about the ngVLA, its progress, scientific outcomes, careers, and its incorporation into the landscape:
\begin{itemize}
\vspace{-6pt}
\item Use of URM students as guides and speakers to provide opportunities for the students to gain experience speaking about their work, raise their visibility, and allow them to serve as role models for members of the audiences.
\vspace{-6pt}
\item Material developed for the road show can also be distributed more broadly through social media, and traditional EPO routes.
\vspace{-6pt}
\item Lesson plans for delivery to local/state schools and after school groups that incorporate radio astronomy and the ngVLA into state standards.
\end{itemize} & BI \& BP & General public; local communities, major towns along the array; K-12, undergrad and grad students; faculty\\

\hline
Opportunities for connections to international colleges and universities to offer post-doctoral exchanges. & BI \& BP & Graduate students\\

\hline
Opportunities for developing materials and resources that are tri-lingual (partners: Mexico and Canada):
\begin{itemize}
\vspace{-6pt}
\item Engage URM students and faculty in the creation of materials.
\end{itemize}& BI \& BP & General public; local communities, K-12, undergrad and grad students; faculty; scientific community\\

\hline
Exploration of opportunities to explore uses and sharable benefits of renewable energy (e.g., solar):
\begin{itemize}
\vspace{-6pt}
\item Partnerships with energy collection and storage industries that may include co-op experiences for URM students and community members.
\end{itemize}& BI \& BP & General public; local communities, K-12, undergrad and grad students; faculty; scientific community\\

\hline
Infrastructure improvement in Magdalena, NM:
\begin{itemize}
\vspace{-6pt}
\item Opportunities for employment and training of local community members.
\end{itemize}& BI \& BP &Local community; workforce\\

\hline
Vocational training for regional students and community members:
\begin{itemize}
\vspace{-6pt}
\item Beta team to conduct remote diagnostics (quick telescope fixes and data back to A-team at HQ).
\vspace{-6pt}
\item Partnerships with regional community colleges.
\end{itemize}& BI \& BP & Undergrad and graduate students at Minority Serving Institutions (MSIs) and other URM-serving institutions\\
\hline
\end{tabular}
\end{center}
\end{table}

\setcounter{table}{0}
\begin{table}[ht]
\caption{Continued}
\begin{center}
\begin{tabular}{|L{8.0cm}|C{1.0cm}|L{3.0cm}|}
\noalign{\smallskip}
\hline

Expanded National Astronomy Consortium (NAC) program:
\begin{itemize}
\vspace{-6pt}
\item Opportunity to develop new cohorts of colleges and universities in ngVLA regions.
\end{itemize} & BP & Undergrad and graduate students at MSIs and other URM-serving institutions\\

\hline
Identification of new outlets for dissemination of technical, engineering, and scientific information:
\begin{itemize}
\vspace{-6pt}
\item National Astronomy Day.
\vspace{-6pt}
\item Engineers Week.
\vspace{-6pt}
\item DragonCon and other similar conferences.
\end{itemize} & BI & General public, scientific community\\

\hline
Greater presence at SACNAS meetings:
\begin{itemize}
\vspace{-6pt}
\item Student and professional recruitment.
\vspace{-6pt}
\item Opportunity to share information about radio astronomy and the ngVLA to a large population of URM students and professionals.
\end{itemize} & BP & undergrad and grad students; faculty; scientific community\\

\hline
Co-op opportunities for students from URM schools to work with the world-class NRAO Central Development and Electronics labs. & BP & undergrad and grad students; faculty; scientific community\\
\hline
\end{tabular}
\end{center}
\end{table}

These ideas represent our commitment to building an ngVLA that is responsible to all members of the scientific and broader communities. We recognize and acknowledge that the ngVLA is indebted to and responsible to taxpayers, local and regional communities, members of underrepresented communities in STEM, K-12 and high education students, postdocs, faculty, scientists, engineers, technical experts, Education and Public Outreach professionals, and all other professions that support the ability to deliver cutting-edge science and technology to the scientific and broader communities.

\section{Summary}
The ngVLA will build on NRAO's existing, well-designed, and implemented, suite of EPO programs, including the creation of new opportunities to educate the general public about the exciting science resulting from ngVLA observations. The ngVLA will make use of, and significantly enhance, the Observatory's flagship National Astronomy Consortium (NAC) and National and International Non-traditional Exchange (NINE) programs, by providing cutting-edge research and engineering opportunities for a new generation of astronomers, engineers, and technicians. In addition to these existing programs, ngVLA will create innovative new BI and BP opportunities that enhance the value of the ngVLA well beyond the scientific community. Importantly, the ngVLA will embrace the NRAO's commitment to providing these opportunities to populations that are unrepresented in STEM fields, and in the field of astronomy in particular.





\end{document}